\documentclass[aps,twocolumn,epsfig]{revtex4}

\usepackage{graphicx}
\usepackage{amsmath,bbm}
\usepackage{amsfonts,amssymb}
\usepackage{color}

\newcommand{\ii}{\mathrm{i}}
\newcommand{\e}{\mathrm{e}}
\newcommand{\op}[1]{{#1}}
\newcommand{\ke}[1]{|{#1}\rangle}
\newcommand{\br}[1]{\langle{#1}|}

\newcommand{\id}{\op I}
\newcommand{\spec}{{\mathcal K}}
\newcommand{\all}{{[N]}}

\newcommand{\cH}{{\mathcal H}}

\begin{document}
\title{Two Notes on Grover's Search: Programming and Discriminating}
\author{Daniel Reitzner$^{1,2}$ and M\'ario Ziman$^{1,3}$} 
\affiliation{
$^1$Institute of Physics, Slovak Academy of Sciences, D\'ubravsk\'a cesta 9, 845 11 Bratislava, Slovakia \\
$^2$Department of Physics, Hunter College of CUNY, 695 Park Avenue, New York, New York 10021, USA\\
$^3$Faculty of Informatics, Masaryk University, Botanick\'a 68a, 60210 Brno, Czech Republic 
}

\begin{abstract}
In this work we address two questions concerning Grover's algorithm. In the first we give an answer to the question how to employ Grover's algorithm for actual search over database. We introduce a quantum model of an unordered phone book (quantum database) with programmable queries to search in the phone book either for a number, or for a name. In the second part we investigate how successful the algorithm can be if the number of elements of the database is not known precisely. This question reduces to analysis of the distinguishability of states 
occurring during Grover's algorithm. We found that using unambiguous discrimination scheme even a seemingly good guess, that is close to the optimal one can 
result in a rather small success rate.
\end{abstract}


\maketitle

\section{Introduction}
\label{sec:intro}

Grover's algorithm \cite{Grover97} is a typical example 
demonstrating the power of quantum computation. It is designed 
to search within an unstructured database of alternatives.
Although it is used in many different applications and algorithms 
of quantum information theory (see \textit{e.g.}~\cite{Ambainis}), we have yet to
succeed to find any explicit example how to use it to search over 
an actual (quantum) database. In the first part of the paper we will 
discuss how to design a (quantum) phone book and 
employ Grover's algorithm to search within either for a name, or for 
a number. In the second part we will investigate how successful the 
algorithm can be if the number of elements of the database is not known (precisely). 
This question is posed as the analysis of the distinguishability of states 
occurring during Grover's algorithm.

Grover's algorithm proves to be quadratically faster 
than any (classical) algorithm performing the task and it was proven to be optimal \cite{BeBeBrVa97} --- there is no quantum (and no classical) algorithm that would do the task faster. The speedup in the algorithm is in the number of oracle calls,
\textit{i.e.}~evaluations of functions
\begin{equation}
\label{eq:class_oracle}
f_{x}(y)=\begin{cases}
1\text{, if $y$ matches $x$,}\\
0\text{ otherwise.}
\end{cases}
\end{equation}
This function evaluates, whether element $x$ has property $y$ (a specific example
is Kronecker's delta which evaluates whether $x=y$). A set of these indexed functions $\{f_x\}_x$
can be represented by the set of paired elements $(x,y)$ which define the database $\cal D$ 
of $N=|{\cal D}|$ elements. In this setting the database search
is conveniently posed as a discrimination among the oracles implementing 
the functions $f_x$. 

For its simplicity we will switch now to the phone book analogy. If we are given the 
task of finding the owner $y$ of the phone number $x$ in the phone book, 
then the phone book is used as an oracle in the following sense
\begin{equation}
\label{eq:phonebook}
f_{x}(y)={\tt Compare}\{x, {\tt Phonebook}(y)\}\,,
\end{equation}
where the function ${\tt Phonebook}(y)$ returns the phone number of owner $y$ and the function
${\tt Compare}(x,x')\equiv f_x(x')$ compares the phone numbers $x$ 
and $x'$, returning $1$ if and only if they match. The database 
search then consists of the identification of the oracle's
input $y$ returning the value $1$.

For the unstructured database (which we can interpret as a search for the owner of a given phone number in the phone book) each
$y$ is equally likely the correct one, \textit{i.e.}~the probability of $(x,y)\in\cal D$ has the same probability for all $y$'s --- if this would not be the case, this prior information would help us search faster in the database. Therefore,
in the classical case the optimal average number of oracle queries identifying
the particular oracle function is $N/2$. At this point the phone book is 
(typically) alphabetically ordered, thus, the Eq.~\eqref{eq:phonebook} 
represents an efficient implementation of the oracle function. However, 
the efficiency of the oracle design is not of interest in the query 
complexity framework. It is assumed to be ``expensive''
in a sense that it requires a lot of resources --- either energy, or time to
return result that is independent of implementation. This being constant justifies
the necessity to count only the number of the times
the oracle is used and the complexity of the algorithm is calculated
in the number of oracle calls.

The quantum algorithm discovered by Grover identifies the oracle 
in $O(\sqrt{N})$ calls, hence, the ability to discriminate quantum 
implementations of different oracle functions requires quadratically
smaller number of queries than in the classical case. Without loss of generality we may assume
that both $x$ and $y$ are indexed from $0$ to $N-1$, and
choose
${\cal D}=\{(x,x): x=0,\dots, N-1\}$. The quantum oracle is 
a quantum analogue of the function (\ref{eq:class_oracle}). In the quantum gate formalism it is implemented as a gate
\begin{equation}
\label{eq:oracle}
\op R_x=\id - 2\ke{x}\br{x},
\end{equation}
with states $\ke x$ forming an orthonormal computational basis of the
$N$-dimensional Hilbert space $\cH_N$. This quantum oracle is a special 
case of a standardly used oracle 
\[
\op V_x\ke y\otimes\ke k=\ke y\otimes\ke{k\oplus f_x(y)},
\]
where $k=0,1$, thus $\op V_x$ acts on the Hilbert space 
$\cH_N\otimes\cH_2$. Initializing the qubit register in the state
$\ke{-}=(\ke{0}-\ke{1})/\sqrt{2}\in\cH_2$ we obtain
$\op V_x\ke y\otimes\ke{-}=(\op{R}_x\ke{y})\otimes\ke{-}$.
From the construction of $R_x$ in Eq.~(\ref{eq:oracle}) follows,
that for $\ke y$ being an element of the chosen computational basis
the states $\ke{y}\otimes\ke{-}$ are eigenvectors of $V_x$ with
eigenvalue either one or minus one (if $y$ is the searched for element).

Each call of the oracle from Eq.~(\ref{eq:oracle}) is in the algorithm 
followed by a unitary operation $\op{G}$ called \emph{inversion about average} 
which acts as
\[
\op{G}=2\ke{\overline{y}}\br{\overline{y}}-\id,
\]
where $\ke{\overline{y}}=\frac{1}{\sqrt{N}}\sum_y \ke{y}$ 
denotes the equal superposition of all computational basis states.

After $m$ repetitive calls of the unitary evolution $\op U_x=\op D\op R_x$ 
the initial query state $\ke{\psi_0}=\ke{\overline{y}}$ evolves into 
\begin{eqnarray}
\label{eq:grover_evolution}
\ke{\psi_m} &=& \sin\frac{(2m+1)\omega}{2}\ke{x}+
\cos\frac{(2m+1)\omega}{2}\ke{\overline{y}_x},
\end{eqnarray}
where $\cos\omega=(N-2)/N$ and $\ke{\overline{y}_x}=
\frac{1}{\sqrt{N-1}}\sum_{y\neq x}\ke{y}$.
We shall call the states $\ke{\psi_m}$ \emph{Grover's states}.
Clearly, if the condition $(2m+1)\omega=\pi$ is met, then 
 $\ke{\psi_m}=\ke{x}$, hence, the search algorithm succeeds --- we
will mark this (in general non-integer) ``number of steps'' with $m_0$.
Strictly speaking, this is possible only for $N=4$, when a single step 
is needed. In all other cases the condition can never be 
exactly reached (for an integer), however, for large $N$ this 
does not cause any problems, as the probability 
\begin{equation*}
\label{eq:successG}
P_G=\sin^2(2m+1)\frac{\omega}{2}
\end{equation*}
will still be sufficiently close to the unity. The optimal number of steps 
scales as $O(\sqrt{N})$ and it was shown \cite{BeBeBrVa97} that
Grover's algorithm is optimal in sense, that it reaches the boundary 
on the number of steps needed to find targeted element $x$. For more details
on Grover's algorithm we refer to any quantum computation textbook, for 
instance \cite{Nielsen+Chuang}.

This paper contains two results on Grover's oracles. In Sec.~\ref{sec:database}
we look closer at the implementation of the oracle and uncover a symmetry
within the ``quantum database''. In Sec.~\ref{sec:discrimination} we evaluate
the quantum search algorithm with unknown size of the database, which reduces
to the discrimination of quantum states appearing during the Grover search algorithm.

\section{Programmable search quantum database}
\label{sec:database}
Let us again switch back to the phone book analogy where Grover's algorithm
searches over now completely unstructured phone book. Not only the numbers of owners
are disordered, but now, for the sake of the argument, let also the owners be
stored randomly in the phone book. Such database $\cal D$ consists of $N$ pairs
$(n,A)$; $n$ will represent phone number and $A$ its owner, ${\cal D}_1$ will be the set of all
the names (persons) in database and ${\cal D}_2$ the set of all the numbers. Let us stress that
both the names and the numbers can be repeated and only pairs of them are
unique.

Let us denote by $\spec_n$ the subset  of people having
the same phone number $n$ and by $\spec_A$ the subset of phone numbers belonging to the
person $A$. Then 
\begin{eqnarray*}
\op{R}_{\spec_n}&=&\id - 2\sum_{A\in \spec_n} \ke{A}\br{A}\,;\\
\op{R}_{\spec_A}&=&\id - 2\sum_{n\in \spec_A} \ke{n}\br{n}\,,
\end{eqnarray*}
are Grover's oracles for searching over the names and the numbers, respectively. 

We now make the key observation for the rest of this section.
It is straightforward to verify that the following identity holds
\begin{equation}
\label{eq:symmetry}
\sum_{n\in\mathcal D_1} \ke{n}\br{n}\otimes\op{R}_{\spec_n}=
\sum_{A\in\mathcal D_2} \op{R}_{\spec_A}\otimes\ke{A}\br{A}\equiv R\,.
\end{equation}
Therefore, the unitary gate
$$
\op{R}=\id\otimes\id -2\sum_{(n,A)\in\cal D} \ke{n}\br{n}\otimes\ke{A}\br{A}\,,
$$
can be understood as the quantum database (oracle) encoding 
the unstructured phone book. 

\begin{figure}
\begin{center}
a) \includegraphics[width=6cm]{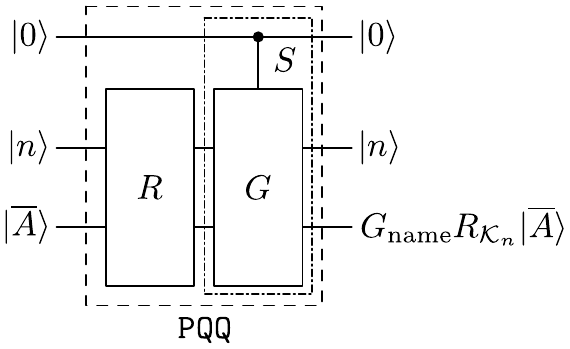} \newline
b) \includegraphics[width=6cm]{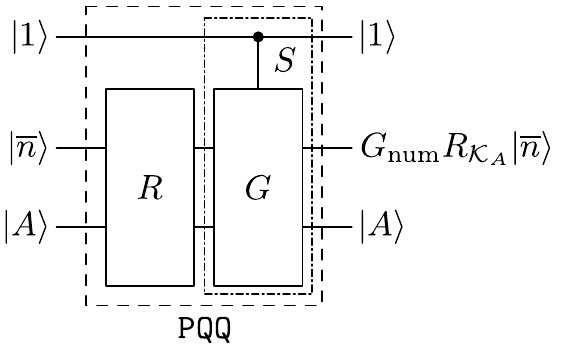}
\caption{The \emph{programmable quantum query} gate  performing one step of the Grover's algorithm over the phone book, when searching a) for the name belonging to the number $\ke n$ or b) for the number belonging to the name $\ke A$. Operation $\op R$ is independent on the type of query and can be considered to be quantum database, while operation $S$ is controlled inversion about average depending on the task performed --- $0$ triggers inversion on the name space, while $1$ triggers inversion on the number space.\label{fig:QQG} }
\end{center}
\end{figure}

Now we will show how Grover's algorithm can be employed to search
over such unstructured phone book. We introduce a programmable quantum 
query gate (\texttt{PQQ} gate) allowing us to run Grover's algorithm to search 
either for a name, or for a phone number in a programmable fashion, 
\textit{i.e.}~the query is represented by the choice of the input state of the 
device and is completely independent of the \texttt{PQQ} gate containing 
the information stored in quantum database $\op{R}$. The \texttt{PQQ} gate 
is illustrated in Fig.~\ref{fig:QQG} and is defined by the following equation
\begin{equation}
{\tt PQQ}=
S_0\otimes(\op{I}\otimes \op{G}_{\rm name})\op{R}
+S_1\otimes (\op{G}_{\rm num}\otimes \op{I})\op{R}\,,
\end{equation}
where 
$\op{G}_{\rm name}=2\ke{\overline{A}}\br{\overline{A}}-\id$, 
$\op{G}_{\rm num}=2\ke{\overline{n}}\br{\overline{n}}-\id$
are the inversions over the respective averages, and $S_j=\ke{j}\br{j}$ 
is a classical (can be made also quantum) switch determining 
whether the name, or the number is going to be searched for, respectively.
Neither the switch, nor the quantum data\-base $\op R$ depend on the 
particular value of the database query. The quantum query (program)
$\ke{1}\otimes\ke{\overline{n}}\otimes\ke{A}$ programs \texttt{PQQ} gate to 
run Grover's search algorithm to identify the phone number 
matching the name $A$. Similarly, the query 
$\ke{0}\otimes\ke{n}\otimes\ke{\overline{A}}$ 
implements Grover's search algorithm to identify 
the name matching the phone number $n$.

In this way we showed, that the programmable oracle, due to the symmetry
(\ref{eq:symmetry}) provides not only a way how to search for the owner
of a phone number, but also the other direction --- how to search for the
phone number of some owner. Both these searches can be made in time
$O(\sqrt{N})$ and, recalling that the database $\op{R}$ is unstructured in
both items, it provides a quadratic speedup in both cases.

Moreover, the
construction can be expanded by an additional type of information,
\textit{e.g.}~mailing address or email, but the overall structure remains the same. Suppose
we have $k$ possible query tasks. The database $\cal D$ consists of $N$ distinct
$k$-tuples $\mathbf x:=(x_0,x_1,\ldots,x_{k-1})$, the oracle (storing the
database) is given as
\begin{equation*}
R=\id^{\otimes k}-\sum_{\mathbf x\in\cal D}\ke{\mathbf x}\br{\mathbf x},
\end{equation*}
where $\ke{\mathbf x}=\ke{x_0}\otimes\ke{x_1}\otimes\cdots\otimes\ke{x_{k-1}}$.
The \texttt{PQQ} gate is then integer-parametrized
\begin{equation}
\label{eq:PQQgen}
{\tt PQQ}=\sum_j S_j\otimes (\id^{\otimes (j-1)}\otimes G_j\otimes \id^{\otimes (k-j)})R
\end{equation}
when performing task $j$ (knowing all other information but $j^\mathrm{th}$); $G_j$ is the corresponding inversion about average on register $j$, $G_j=2\ke{\bar y_j}\br{\bar y_j}-\id$.

Higher degree of free parameters allows also a wider variety of problems than the one mentioned above which just serves to fill in the information $j$ while the rest is known. In general we can be given a smaller subset of parameters characterizing the element
we want to find in the database (\textit{e.g.}~knowing the phone number and email, we might want to find the name and address of the owner).
This general case does not differ much from the previously discussed cases. The initial state is prepared as the equal superposition over the basis states of all unknown subspaces and as a given choice on the subspaces where the information about the searched element is known. The \texttt{PQQ} is then similar to Eq.~(\ref{eq:PQQgen}) with $j$ indexing the possible types of searches we might want to perform --- the corresponding term in Eq.~(\ref{eq:PQQgen}) for given $j$ will be then $S_j$ tensored with operator having identity operator $I$ on all the positions the information is known and respective $G$ on all the positions the information is unknown to us.

Using the oracle point of view the construction and/or performance of the quantum database $\op R$ is not an issue, however, from application point of view this question (especially the performance) is of high relevance. Here we have addresed only one implementation problem: the actual design of quantum database. The questions related to writing, or deleteing entries from database we left untouched.

\section{Grover's search with unknown size of the database.}
\label{sec:discrimination}

The size of the Hilbert space we search for might be unknown, or not known precisely.
It is generally a difficult problem to decide what the size of the Hilbert space is, especially when
it might be rather large \cite{scarani,wolf}. Having Grover's states as resources and being able to
choose only the number of steps $m$ after which Grover's algorithm stops we
might therefore not know how close we are to the optimal number of steps and we
want to know how reliable our results will be. This question 
can be recast as a discrimination of quantum states produced by Grover's oracles 
after $m$ uses, hence, the question is how distinguishable the 
states from Eq.~\eqref{eq:grover_evolution} are. We will investigate two
extreme variations of the problem: the minimum-error discrimination optimizing
the average success rate of our conclusions and the unambiguous discrimination
allowing for error-free conclusions while tolerating inconclusive outcomes.

Our goal is
to find a final measurement optimizing the associated success rates
while keeping the rest of Grover's algorithm unchanged. Let us stress 
this problem is different from discrimination of Grover's oracles, where one
is allowed to design also the test state and to employ ancillary systems
and devices in order to optimize the success rates. In Ref.~\cite{ChKiTaSaTw07}
some results on unambiguous discrimination of Grover oracles are given, stating
that the unambiguous discrimination of Grover's oracles is always possible. The exact protocols achieving
perfect (error-free) discrimination of Grover's oracles were investigated in \cite{WuDu08},
where it was shown that in order to achieve such goal the number of queries scales
as $N-\sqrt{N}$ with the size of the database. It achieves better scaling than any
classical algorithm requiring at least $N-1$ oracle calls. However, the quadratic speed-up
is in this case lost. As far as we know, the oracle discrimination problem is still open.
Grover's algorithm provides \cite{Za1999} an asymptotic solution to minimum-error
case quantifying the number of queries needed for vanishingly small error. 

\subsection{Symmetry of Grover's oracles}
Before we proceed let us note that Grover's oracles $U_{x}^m$
respect the following symmetry
\[
\op T\op U^m_{x-1}\op T^\dagger=\op U^m_x\;,
\]
where 
\begin{equation}\label{eq:T}
T=\sum_x \ke{(x+1){\rm mod}N}\br{x}
\end{equation}
is the shift operator with $T^N=I$. The eigenvalues of $T$ 
are $\lambda_a=\exp(\ii2\pi a/N)$ for $a=0,\dots,N-1$ and 
corresponding eigenvectors are
\[
\ke{\gamma_a}=\frac{1}{\sqrt{N}}\sum_{y=0}^{N-1}\e^{-\ii\frac{2\pi ay}{N}}\ke y.
\]
This symmetry feature has a favorable mathematical consequence. If we 
take the initial state of equal superposition $\ke{\psi_0}$, which is invariant under the action
of $T$, \textit{i.e.}~$T\ke{\psi_0}=\ke{\psi_0}$, then the output states 
$\ke{\psi_x(m)}=U_x^m\ke{\psi_0}$  will respect the same symmetry as the unitary matrices $U_x^m$. 
In this way, for each step $m$ the Grover states $\ke{\psi_x(m)}=
U_x^m\ke{\psi_0}$ form a family of symmetric states
satisfying the relation $\ke{\psi_x(m)}=T^x\ke{\psi_0(m)}$, where
$\ke{\psi_0(m)}=U_{x=0}^m\ke{\psi_0}$. This reduces our discrimination 
problems to discrimination of symmetric 
states $\{\ke{\psi_x(m)}\}_{x}$ being the set of potential 
output states after $m$ steps of Grover's algorithm. 

\subsection{Unambiguous discrimination}
Let us start with the case of unambiguous discrimination \cite{sedlak}. In this case, the conclusions made are certain, hence, the algorithm is exact although it requires an inconclusive result. In Ref.~\cite{ChBa98} a theory of unambiguous discrimination of (pure) symmetric states is described. In particular, if we are given a set of $N$ pure symmetric states $\ke{\phi_x}=T^x\ke{\phi_0}$ for some unitary operator $T$ (such that $T^N=I$), then using the result of Ref.~\cite{ChBa98} we can evaluate the upper bound on probability of success in unambiguous discrimination as
\begin{equation}
\label{eq:unambiguous_prob}
P_{\rm suc}\leq N\min_a |\langle{\gamma_a}|{\phi_0}\rangle|^2\,,
\end{equation}
where $\ke{\phi_0}$ is the test state and $\ke{\gamma_a}$ 
are the eigenvectors of $T$. 

In our case $T$ is given by Eq.~\eqref{eq:T} and we are to discriminate the states $\{T^x\ke{\psi_0(m)}\}_x$
given $\ke{\phi_0(m)}=U_0^m\ke{\gamma_0}$. We find
\begin{eqnarray}
\nonumber
P_{\rm suc}(m)&\leq &
N\min_a |\br{\gamma_a}U_0^m\ke{\gamma_0}|^2\equiv\Gamma_0(m) \,.
\end{eqnarray}
Let us denoty by $\ke{\overline{\gamma}}=
\frac{1}{\sqrt{N}}\sum_a \ke{\gamma_a}=\ke{0}$ and by 
$\ke{\overline{\gamma}_0}=\frac{1}{\sqrt{N-1}}\sum_{a\neq 0}\ke{\gamma_a}$. 
Then a single step of Grover's algorithm can be expressed as  
\begin{eqnarray}
\nonumber
U_0&=&2\ke{\gamma_0}\br{\gamma_0}+2\ke{\overline{\gamma}}\br{\overline{\gamma}}
-I-\frac{4}{\sqrt{N}}\ke{\gamma}\br{\overline{\gamma}}\\
&=&(I_0-I)+\left[(1-\frac{2}{N})I_0 -i\frac{2\sqrt{N-1}}{N}Y_0\right]\,,
\end{eqnarray}
where $I_0$, $Y_0$ are Pauli operators defined on two-dimensional 
subspace $\cH_0$ spanned by the vectors $\ke{\gamma_0}$, $\ke{\overline{\gamma}_0}$,
thus, $I_0=\ke{\gamma_0}\br{\gamma_0}+\ke{\overline{\gamma}_0}
\br{\overline{\gamma}_0}$, $Y_0=-i\ke{\gamma_0}\br{\overline{\gamma}_0}+i\ke{\overline{\gamma}_0}\br{\gamma_0}$, and $I-I_0$ is the projector onto the 
orthogonal subspace $\cH_0^\perp$. As in the original Grover's algorithm we
define the angle $\omega$ via the identity $\cos\omega=1-2/N$. Then
$$
U_0^m=(I_0-I)+e^{-im\omega Y_0}\,.
$$

\begin{figure}
\begin{center}
\includegraphics[width=8cm]{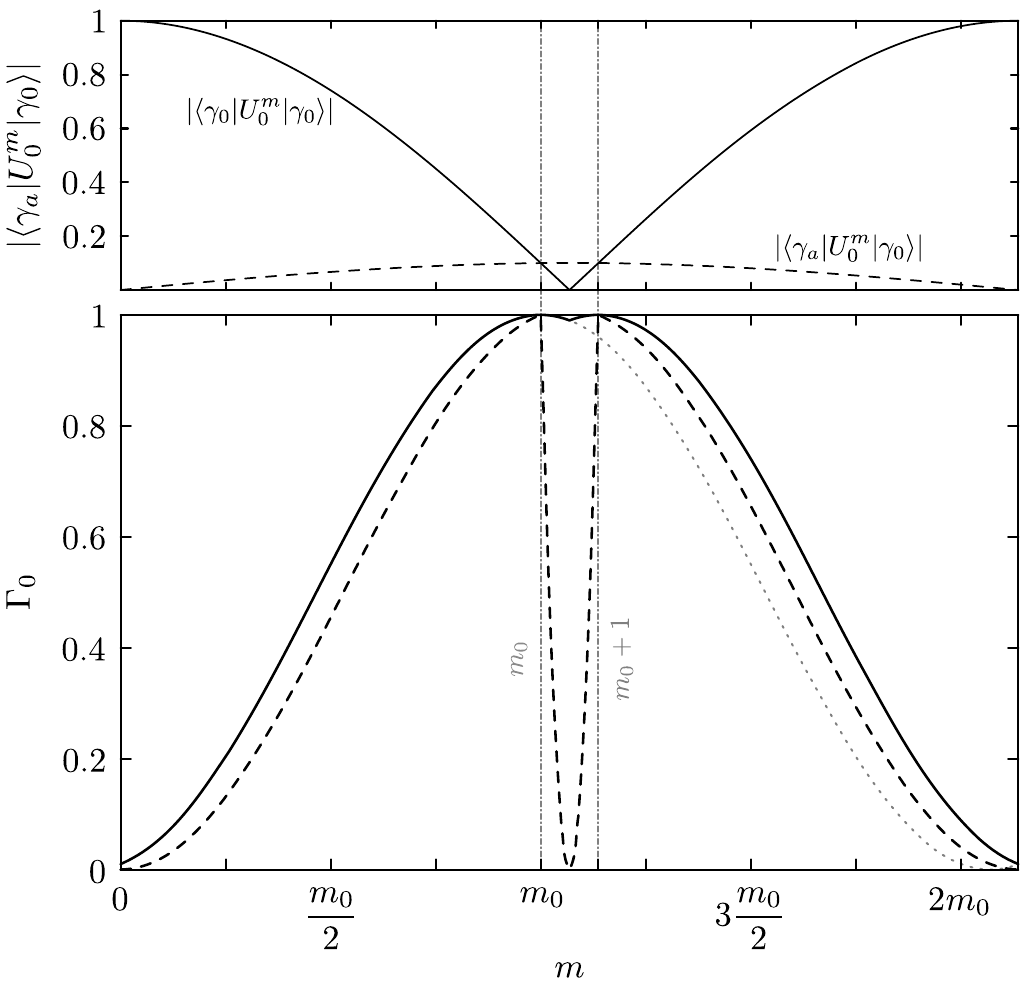}
\caption{\label{fig:graph} 
Illustration (with exaggerated differences --- small size of database with $N=100$) of the bound on success probability 
for discrimination of Grover's oracles. The upper plot depicts the terms in $\Gamma_0$ over which we minimize, while the lower plot shows
the success probabilities for the different discrimination schemes --- unambiguous discrimination (dashed line) has a dip between optimal number of steps $m_0$ and $m_0+1$ that can lead to unsuccessful discrimination. The minimum-error discrimination (solid line) does not suffer this problem and up to the point $m_0$ copies the usual Grover's success probability.
}
\end{center}
\end{figure}

Using the above form of $U_0^m$ we find
\begin{eqnarray}\label{eq:sucgamma} 
\Gamma_0(m)
=\min\left\{|\cos\omega m|, \frac{|\sin\omega m|}{\sqrt{N-1}}\right\}\,.
\end{eqnarray}
The minimized elements of this function (which is the upper bound on the success probability for unambiguous discrimination) is plotted in Fig.~\ref{fig:graph} (upper plot). Since $|\cos m\omega|$ and $|\sin m\omega|$ have exactly opposite monotonicity, it follows that the maximal value (with respect to $m$) is achieved when they coincide, \textit{i.e.}~when $|\cos m\omega|=|\sin m\omega|/\sqrt{N-1}$. This condition gives us two solutions $m_0$ and $m_0+1$ when the perfect discrimination is possible as the Grover's states become orthogonal. Also it is not surprising that (in the limit of $N\to\infty$) the success probability approaches 1 for the number of calls coinciding with the number of calls needed in Grover's search. Indeed, at this point different oracles lead to mutually orthogonal quantum states.

Relating to the question we answer in this paper, knowing the length of the database (the size of the Hilbert space) only approximately, an interesting observation is at hand. If our chosen number of steps $m$ will be the closest integer larger than the optimal number $m_0$ (we recall, that $m_0$ is integer only for $N=4$) but smaller than $m_0+1$ the unambiguous discrimination scheme can fail as the minimal term in Eq.~(\ref{eq:sucgamma}) will be the cosine term going to zero. If $\omega m$ is close to $\pi/2$ --- this happens when $m=m_0+1/2$ --- by Eq.~(\ref{eq:sucgamma}) the success probability will be bounded from above by $0$ and the search will be unsuccessful as Grover's states in this case are linearly dependent. This can however exactly happen only for $N=2$ but one can get very close to this point for large $N$ as well and the success probability can be very small, as after $m_0$ it drops fast towards zero --- see the lower plot of Fig.~\ref{fig:graph}.

\subsection{Minimum-error discrimination}

In the case of minimum-error discrimination the results from Ref.~\cite{Holevo73} provide necessary and sufficient conditions for discriminating states, while in Ref.~\cite{BaCr09} specific results on the discrimination of states are provided. Minimum-error discrimination of pure symmetric states was addressed in Ref.~\cite{BaKuMoHi97}, where the optimal success probability $P_M$ was shown to be
\begin{equation}
\label{eq:success_minerr}
P_M=|\br{\psi_0(m)}\Omega^{-1/2}\ke{\psi_0(m)}|^2
\end{equation}
with
\begin{align*}
\Omega &=\sum_{x\in\all}\ke{\psi_x(m)}\br{\psi_x(m)}\\
&=N\cos^2 m\omega\ke{\overline y}\br{\overline y}+\frac{N}{N-1}\sin^2 m\omega(\id-\ke{\overline y}\br{\overline y}).
\end{align*}
Since $|\langle\psi_0(m)\ke{\overline y}|^2=\cos^2 m\omega$, from Eq.~(\ref{eq:success_minerr}) we find
\[
P_M = \left|\frac{1}{\sqrt{N}}\cos m\omega+\sqrt{\frac{N-1}{N}}\sin m\omega\right|^2.
\]
From this equation we obtain (see also Fig.~\ref{fig:graph})
\[
P_M=\begin{cases}
\sin^2(2m+1)\omega/2\text{, for $m\leq m_0+1/2$,}\\
\sin^2(2m-1)\omega/2\text{, for $m\geq m_0+1/2$.}\\
\end{cases}
\]
Again we may notice perfect discrimination ($P_M=1$) not only at $m=m_0$ but also at $m=m_0+1$ when the states would be orthogonal and the minimum-error discrimination coincides with the unambiguous discrimination. For choice of $m$ smaller than $m_0+1/2$ the success probability copies that of the usual Grover's search, and for $m$ larger it becomes slightly advantageous.
If the choice of $m$ falls in the region $[m_0,m_0+1]$, in contrast to the unambiguous discrimination scheme where the probability drops towards zero, we do not have any considerable drop in probability showing that minimum-error discrimination is in this sense superior to the unambiguous discrimination scheme. Furthermore, considering only integer $m$, none of the discrimination schemes can be perfect. 

\section{Conclusion}

We have introduced the concept of programmable search 
database (see Fig.~\ref{fig:QQG}) employing (in a programmable way) 
Grover's oracles to search over an unstructured databases (like phone book). It
enables us to choose query (either name, or phone number) and search for its
complement (phone number, or name, respectively) from the unstructured data\-base. 
Because of the symmetry of the pro\-gram\-ma\-ble search database for any query the 
complexity is the same as for Grover's algorithm but offers a lot of flexibility.
Moreover, this construction works also for higher degree of searchable items (like mailing address, email, etc.).
We believe this note clarifies how the Grover algorithm might actually be used for searching a quantum database, especially
with more degrees of freedom within which one might want to search.
Although we have not addressed the question of how the database would be physically constructed, this note provides an outlook on what one should consider --- the symmetry of the oracle, if implemented, would make the search more universal.

In the second note we have discussed the performance of Grover's search algorithm when the size of the database is unknown, but the resources (probe state and oracles) are available at user's disposal. We have found that the measurement point has to be chosen carefully (even if the guess is almost precise), as in a small range between the points of perfect discrimination, the success probability can drop significantly (see Fig.~\ref{fig:graph}). This feature holds for unambiguous approach and therefore minimum error might be favored more if the size of the database is not known exactly. Minimum-error discrimination seems to be more practicable as it not only overcomes the pit near $m_0+1/2$ but it also works in the presence of small errors. Moreover, it might be applied more easily, as the bound for unambiguous discrimination can be hard to reach. Finally, as for the number of steps smaller than $m_0$ it copies the usual success probability for Grover's search we see, that the measurement in the computational basis performs the minimum error discrimination. The unambiguous discrimination, although very interesting from the theoretical point of view, is to large extent impractical.

The two presented notes cover only a small set of directions of interest where only partial results are known. For example we still do not know what an actual realization might look like --- quite possibly it will be a subroutine in a larger algorithmic application \cite{Ambainis}. Other interesting directions to pursue are geometric analysis of Grover's search \cite{mancini} or quantum searches under decoherence \cite{SBW,RegevSchiff}.


\section*{Acknowledgment}
This work was supported by projects APVV-0808-12 (QIMABOS)  
and COST Action MP1006. M.Z.~acknowledges the support of 
RAQUEL and GA\v CR project P202/12/1142. D.R.~acknowledges
the support of Fulbright Visiting Scholar Program.

\end{document}